\documentclass[preprint]{aastex}

\slugcomment{KSUPT-01/1 \hspace{0.5truecm} February 2001}

\shortauthors{Podariu et al.}
\shorttitle{Binned CMB Anisotropy Power}

\begin{document}

\title{Binned Cosmic Microwave Background Anisotropy \\
       Power Spectra: Peak Location}

\author{Silviu Podariu\altaffilmark{1},
        Tarun Souradeep\altaffilmark{1,2},
        J. Richard Gott, III\altaffilmark{3},
        Bharat Ratra\altaffilmark{1}, \\
        and
        Michael S. Vogeley\altaffilmark{4}
       }
\altaffiltext{1}{Department of Physics, Kansas State University, Manhattan, 
                 KS 66506.}
\altaffiltext{2}{Current address: IUCAA, Post Bag 4, Ganeshkhind, Pune 411007,
                 India.}
\altaffiltext{3}{Princeton University Observatory, Princeton, NJ 08544.}
\altaffiltext{4}{Department of Physics, Drexel University, Philadelphia, 
                 PA 19104.}

\begin{abstract}
We use weighted mean and median statistics techniques to combine
individual cosmic microwave background (CMB) anisotropy detections and
determine binned, multipole-space, CMB anisotropy power spectra. The
resultant power spectra are peaked. The derived weighted-mean CMB
anisotropy power spectrum is not a good representation of the
individual measurements in a number of multipole-space bins, if the
CMB anisotropy is Gaussian and correlations between individual
measurements are small. This could mean that some observational error
bars are underestimated, possibly as a consequence of undetected
systematic effects. Discarding the most discrepant 5\% of the
measurements alleviates but does not completely resolve this
problem. The median-statistics power spectrum of this culled data set
is not as constraining as the weighted-mean power
spectrum. Nevertheless it indicates that there is more power at
multipoles $\ell \sim 150-250$ than is expected in an open cold dark
matter (CDM) model, and it is more consistent with a flat CDM model.
Unlike the weighted-mean power spectrum, the median-statistics power
spectrum at $\ell \sim 400-500$ does not exclude a second peak in the
flat CDM model.
\end{abstract}

\keywords{cosmic microwave background---cosmology: observation---methods:
  statistical---methods: data analysis---large-scale structure of the 
  universe}

\section{Introduction} 

Current observational data favors low-density cosmogonies. The simplest 
low-density models have either flat spatial 
hypersurfaces and a constant or time-variable cosmological ``constant" 
$\Lambda$ (see, e.g., Peebles 1984; Peebles \& Ratra 1988; Sahni \& 
Starobinsky 2000; Steinhardt 1999; Carroll 2000; Bin\'etruy 2000), or 
open spatial hypersurfaces and no $\Lambda$ (see, e.g., Gott 1982, 1997;
Ratra \& Peebles 1994, 1995; Kamionkowski et al. 1994; G\'orski et al. 1998).
Typically the CMB anisotropy power spectrum is predicted to peak at larger
multipole moment $\ell$ (smaller angular scale) in the open case than in the 
flat model. This difference in the flat and open model power spectra makes it
possible for CMB anisotropy measurements to distinguish between these models.
See, e.g., Barreiro (2000), Rocha (1999), Page (1999), and Gawiser \& Silk 
(2000) for recent reviews of the field.

Until very recently no single CMB anisotropy experiment achieved detections
over a wide enough range of $\ell$ space to allow this cosmological test to 
be performed with data from a single experiment\footnote{
The recent BOOMERanG 1998 (de Bernardis et al. 2000) and MAXIMA-1 (Hanany et 
al. 2000) detections of CMB anisotropy do cover a wide enough range of 
$\ell$ space for the test. By including 2 $\sigma$ upper limits 
on CMB anisotropy (see discussion below), the MAT 1998 (Miller et al. 1999),
Viper (Peterson et al. 2000), and BOOMERanG 1997 (Mauskopf et al. 2000) 
observations also cover a wide enough range of $\ell$ space.}.  
As a result the test has usually been performed as a goodness-of-fit 
($\chi^2$) comparison of CMB anisotropy model predictions and observations
(Ganga, Ratra, \& Sugiyama 1996)\footnote{
See Knox \& Page (2000) for an alternate approach to this test.}. 
In this implementation the test favors a 
flat model over an open one (see, e.g., Dodelson \& Knox 2000; Tegmark \&
Zaldarriaga 2000; Le Dour et al. 2000; Lange et al. 2001; Balbi et al. 2000,
2001; Amendola 2001). However, this $\chi^2$ technique has a number of 
significant limitations, so these results must be viewed as tentative 
(see discussion in Ratra et al. 1999b).

Specifically, a $\chi^2$ comparison does not use the complete data from an
experiment. Rather, it uses one (or a few) number(s) (the amplitude of a 
chosen model CMB power spectrum, typically the flat bandpower spectrum) with 
error bars, as a summary of the complete data of the experiment. However, this 
amplitude (and error bars) is model dependent. This effect is typically 
$\sim 10\%$ for a data set with a good detection (see, e.g., Ganga
et al. 1997a, 1998; Ratra et al. 1999a), but is not accounted for
in the $\chi^2$ comparison. More significantly, the observational error
bars are derived from non-Gaussian posterior probability
density distribution functions and are thus fairly asymmetric. Since the 
$\chi^2$ technique assumes symmetric (Gaussian) error bars, the observational 
error bars must be symmetrized (``Gaussianized") when this technique is
used. This is an arbitrary procedure, and Gaussianizing in different ways
leads to different reduced $\chi^2$ values (Ganga et al. 1996). This means that
the $\chi^2$ technique can only provide qualitative results. Nevertheless, 
it is useful, having provided, from a combined analysis of all available
early observational results (Ratra et al. 1997), qualitative evidence for
more CMB anisotropy power on smaller scales than on larger scales (Ganga et
al. 1996), consistent with later observations by single experiments (see,
e.g., Netterfield et al. 1997; de Oliveira-Costa et al. 1998; Coble et al.
1999; Miller et al. 1999; Peterson et al. 2000; de Bernardis et al. 2000; 
Hanany et al. 2000).   

Given that current observational error bars are asymmetric (i.e., non-Gaussian),
robust results can only be derived from a complete maximum likelihood 
analysis of a large collection of observational data, using realistic 
model CMB anisotropy power spectra. This is a very time-consuming approach
and so has been applied to only a few data sets (see, e.g., Bunn \& Sugiyama
1995; G\'orski et al. 1995; Ganga et al. 1997a, 1997b;
Stompor 1997; Ratra et al. 1998; Rocha et al. 1999). Ganga et al. (1997a) 
generalized the maximum likelihood technique to account for systematic errors
(e.g., beamwidth uncertainty, calibration uncertainty). Analyzes using this 
generalized technique have led to significantly revised observational results. 

The weighted mean and median statistics techniques are summarized in the next 
section. In $\S$3 we use these techniques to determine binned CMB anisotropy
power spectra using all CMB anisotropy measurements and in $\S$4 we consider
a culled collection of good measurements. We conclude in $\S$5.

\section{Binned CMB Anisotropy Power Spectra: Techniques}

While tight constraints on cosmological parameters will be very valuable, it
is also of interest to determine the shape of the observed CMB anisotropy
spectrum in a model independent fashion. Since the error bars for experiments 
to date are large, a tight determination of the CMB anisotropy spectrum is 
possible only if all available data are used. This approach has been pioneered 
by Page (1997, 1999). 

The data we use in this paper are listed in Table 1 and plotted in Figure 1. 
We consider all detections at $\ell < 1000$ available as of early November 2000.
While we only use detections, defined as those results where the peak of the
likelihood function is at least 2 $\sigma$ away from 0 $\mu$K, in our 
analyses here (and these are listed in Table 1), Figure 1 also shows 2 $\sigma$
upper limits. Typically, the bandtemperature ($\delta T_\ell$) values listed in
Table 1 have been derived assuming a flat bandpower spectrum and account for 
known systematic uncertainties (although in most cases only in an approximate
manner). Since the flat bandpower spectrum is a more accurate representation
of the true spectrum for narrower (in $\ell$) window functions, we use 
observational results for the narrowest windows available. Table 1 lists 142 
detections.

To determine the observed CMB anisotropy spectrum, Page (1997, 1999) binned
the detections into equally spaced logarithmic bins in $\ell$-space, based
on the value of $\ell_{\rm e}$, the effective multipole of the experiment's
window function $W_\ell$. Here $\ell_{\rm e} = I(\ell W_\ell)/I(W_\ell)$, 
where $I(W_\ell) = \sum_{\ell = 2}^\infty (\ell + 0.5) 
W_\ell/\{\ell(\ell + 1)\}$. As discussed below, we instead choose to adjust the 
$\ell$-space widths of the bins so as to have approximately the same 
number of measurements in each bin. Page then Gaussianized the 
measurements\footnote{
Bond, Jaffe, \& Knox (2000) discuss a more accurate approximation that
retains some of the non-Gaussianity.}
by defining the error of a measurement, $\sigma$, to be half the difference
between the upper and lower 1 $\sigma$ values of $\delta T_\ell$ for the 
measurement. The standard expression for the weighted mean in bin $B$ is
\begin{equation}
  \delta T_\ell^B = {\sum_{i=1}^{N_B} (\delta T_\ell)_i / \sigma_i{}^2
      \over \sum_{i=1}^{N_B} 1 / \sigma_i{}^2} ,
\end{equation}
where $i = 1, 2, \dots N_B$ indexes the $N_B$ measurements in the bin with
bandtemperature central values $(\delta T_\ell)_i$ (where the likelihood
is at a maximum) and Gaussianized errors $\sigma_i$. The (internal) error
estimate for each bin is
\begin{equation}
  \sigma^B = \left(\sum_{i=1}^{N_B} 1 / \sigma_i{}^2\right)^{-1/2} .
\end{equation} 

To plot the observed CMB anisotropy power spectrum, Page places 
$\delta T_\ell^B$ at the arithmetic mean of the $\ell_{\rm e}$ values of the 
measurements that lie in the bin. We choose instead to use the weighted 
mean of the $\ell_{\rm e}$ values,
\begin{equation}
  \ell_{\rm e}^B = {\sum_{i=1}^{N_B} (\ell_{\rm e})_i / \sigma_i{}^2
      \over \sum_{i=1}^{N_B} 1 / \sigma_i{}^2} .
\end{equation}

Since the weighted mean technique assumes Gaussian errors, one may compute a 
goodness-of-fit parameter, $\chi^2_B$, for each bin,
\begin{equation}
  \chi^2_B = {1 \over N_B - 1} \sum_{i=1}^{N_B} {\left((\delta T_\ell)_i 
      - \delta T_\ell^B\right)^2 \over \sigma_i{}^2} .
\end{equation}
$\chi_B$ has expected value unity with error $1/\sqrt{2(N_B - 1)}$, so
\begin{equation}
  N^B_\sigma = |\chi_B - 1| \sqrt{2(N_B - 1)}
\end{equation} 
is the number of standard deviations that $\chi_B$ deviates from unity.
A large value of $N^B_\sigma$ could indicate the presence of 
unaccounted-for systematic uncertainties, the invalidity of the Gaussian
assumption, or the presence of significant correlations between the 
measurements.

An alternate method for deriving the observed CMB anisotropy spectrum
is the median statistics approach developed by Gott et
al. (2001). Here one does not assume that the measurement errors are
Gaussian, or even that the magnitudes of the errors are known. One
assumes only that the measurements are independent and free of
systematic errors. The technique is discussed in detail in Gott et
al. (2001).  In brief, for each bin in $\ell$-space we construct a
likelihood function for the true median of the binned
measurements\footnote{ As described in Gott et al. (2001), this is a
histogram that gives the relative probability of obtaining $\delta
T_\ell$. In principle, the highest $\delta T_\ell$-space bin in this
histogram extends to $\infty$ $\mu$K; in practice we have picked the
width of this bin to be the same as the width of the lowest $\delta
T_\ell$-space bin in the histogram. This prescription controls the
large $\delta T_\ell$ divergence when we integrate over the likelihood
function with a logarithmic prior (see below). We have ensured that
each $\ell$-space bin contains a sufficient number of measurements so
that the upper 2 $\sigma$ limit on $\delta T_\ell^B$ always lies below
the highest $\delta T_\ell$-space bin. Therefore the upper 2 $\sigma$
limit on $\delta T_\ell^B$ is insensitive to the procedure used to
control the large-$\delta T_\ell$ divergence.}  and then integrate
over this with a logarithmic prior to determine the error bars for the
bin.  We use a logarithmic prior since $\delta T_\ell$ is positive
definite (see discussion in Gott et al. 2001). We have checked that a
linear prior leads to qualitatively similar conclusions, with
generally only small quantitative differences. In this case $\delta
T_\ell^B$ is the median measurement, which is defined to be the mean
of the two central measurements if the bin contains an even number of
measurements. Since we determine limits on $\delta T_\ell^B$ in each
$\ell$-space bin by integrating the likelihood function for the bin,
for accurate 2 $\sigma$ limits we need to ensure that enough
measurements lie in each bin. Consequentially, in contrast to Page
(1997, 1999), we adjust the widths of the bins so that each of them
contain about the same number of measurements, with precise bin
membership determined by where breaks occur in the $\ell$-space
distribution of the measurements of Table 1. And, instead of plotting
$\delta T_\ell^B$ at the weighted mean of the $\ell_{\rm e}$ values of
the measurements in the bin, in the median-statistics case we use the
median of the $\ell_{\rm e}$ values in the bin. Since the median
statistics technique ignores the individual measurement errors there
is no obvious way of checking the internal goodness of fit of the
derived median-statistics CMB anisotropy power spectrum.

\section{Binned CMB Anisotropy Power Spectra Using All Measurements}

We first consider four different binnings, with about 9, 11, 13, and
16 measurements in each $\ell$-space bin. Tables 2 and 3 list results
from the weighted-mean and median-statistics analyses. In the
median-statistics case, the logarithmic prior makes the integral of
the likelihood function diverge at large and at small $\delta
T_\ell$. We discuss our prescription for controlling the large-$\delta
T_\ell$ divergence in footnote 7. To control the small-$\delta T_\ell$
divergence we cut off the integral at small $\delta T_\ell$, with the
cutoff chosen so that the lower 2 $\sigma$ limit is insensitive to it
(see footnote b of Table 3 for the numerical value of the cutoff
used). Figures 2 and 3 show the weighted-mean and median-statistics
observed CMB anisotropy power spectra.

The weighted mean analysis (Figure 2 and Table 2) results in tight
constraints on the observed CMB anisotropy power spectrum, and clearly
establishes that it has a peak. For three of the four binnings used
this is a rather broad peak\footnote{ We define the interval in which
the peak lies by the $\ell_{\rm e}^B$ values of those bins whose
amplitudes are within $90\%$ of the maximum $\delta T_\ell^B$ value,
for the weighted-mean central value and the $\pm$1 $\sigma$ and $\pm$2
$\sigma$ limits.}, and lies in the interval $\ell \sim 170-240$ (for 9
measurements per bin, Figure 2$a$), $\ell \sim 180-210$ (for 11
measurements per bin, Figure 2$b$), and $\ell \sim 190-240$ (for 13
measurements per bin, Figure 2$c$).  For the case of 16 measurements
per bin the peak consists of a single bin at $\ell \sim 210$. As
expected, the CMB anisotropy power spectrum has less scatter as a
function of $\ell$ when the number of bins is decreased.

We have not considered upper limits in our analyses. From Figure 2 we
see that at least four of these are quite constraining and if correct
they could significantly affect the shape of the observed power
spectrum when accounted for.  These are the $\ell = 15$, 16, and 20
DMR upper limits and the $\ell = 138.7$ MAX5 $\mu$ Pegasi upper
limit. The DMR results are derived from the galactic frame maps and
ignore the Galactic emission correction (G\'orski 1997; G\'orski et
al. 1998). Therefore they do not account for the full uncertainty in
the DMR data. Our analyses also ignore the correlations between the
different DMR $\ell$-space results. To derive the MAX5 $\mu$ Pegasi
upper limit Ganga et al. (1998) marginalized over a possible dust
contaminant signal. Ganga et al. (1998) concluded that the MAX5 $\mu$
Pegasi upper limit was not inconsistent with the other MAX4 and MAX5
results they studied\footnote{ Note from Figure 1 that in the MAX5
$\mu$ Pegasi upper limit region of $\ell$-space there are a number of
detections with low $\delta T_\ell$.  These are responsible for the
prominent drop in this bin in Figure 2, especially in panels $b$ and
$c$.}.  It is quite possible that knowledge of the dust contaminant
signal is less than adequate for the purpose of extracting a robust
constraint on the CMB data in this case. It is troubling that three
published 2 $\sigma$ upper limits lie at or below the central weighted
mean values derived from published detection results.

More worrisome are the large values of $N^B_\sigma$ (the dashed line
in each panel of Figure 2 and the last column of Table 2) for some of
the bins. For a Gaussian CMB anisotropy, $N^B_\sigma$ is a measure of
how well the weighted mean and derived bin error bar represents the
measurements that lie in the bin.  This is larger than 2 (i.e.,
$\chi_B$ is more than 2 $\sigma$ away from what is expected for a
Gaussian distribution) for 5 of 16 bins for the 9 measurements per bin
case (bin numbers 1, 5, 8, 9, and 15), for 4 of 13 bins for the 11
measurements per bin case (bin numbers 1, 4, 8, and 12, and for bin
number 11 $N^B_\sigma = 1.9$), for 3 of 11 bins for the 13
measurements per bin case (bin numbers 1, 7, and 10), and for 3 of 9
bins for the 16 measurements per bin case (bin numbers 1, 5, and
8). Note that $N^B_\sigma < 1.6$ for $\ell$-space bins in the peak
intervals discussed above (see Table 2).

Since at least two-thirds of the bins have small $N^B_\sigma$ it is
unlikely that the CMB anisotropy is non-Gaussian. The large values of
$N^B_\sigma$ are more likely caused by unaccounted-for foreground
contamination, or other effects that lead to underestimated error bars
on some of the measurements, and our neglect of correlations between
some of the measurements. Since the second effect is thought to be
small, the first effect is probably the dominant one. It is important
to note that this inconsistency implies that constraints on
cosmological parameters derived from $\chi^2$ comparisons of multiple
CMB anisotropy observations and model predictions must be interpreted
with care.

The median statistics technique does not make use of the error bars on
the measurements. Therefore it is ideally suited for an analysis of
this combination of CMB anisotropy data. The median statistics
analyses result in somewhat weaker constraints on the observed CMB
anisotropy detections power spectrum (Figure 3 and Table 3), but still
clearly establish that it has a peak. The median-statistics peak
interval is slightly broader and extends to slightly smaller angular
scales than the weighted-mean peak interval. The median-statistics
peak lies in the interval $\ell \sim 160-250$ (for 9 measurements per
bin, Figure 3$a$), $\ell \sim 150-260$ (for 11 measurements per bin,
Figure 3$b$), $\ell \sim 180-240$ (for 13 measurements per bin, Figure
3$c$), and $\ell \sim 200-290$ (for 16 measurements per bin, Figure
3$d$). It is significant that the median-statistics constraints on
$\delta T_\ell^B$ are significantly weaker than the weighted-mean
constraints on $\delta T_\ell^B$ for those bins with large
$N^B_\sigma$ values from the weighted mean analysis. We note that,
although the median-statistics constraints are weaker, the
median-statistics power spectra of detections are also inconsistent
with the three 2 $\sigma$ upper limits that are a problem for the
weighted-mean power spectra.

\section{Binned CMB Anisotropy Power Spectra Using ``Good" Measurements Only}

In the weighted mean analyses of the previous section we found that a
number of $\ell$-space bins had large values of $N^B_\sigma$. We
argued that these large $N^B_\sigma$ values were likely the
consequence of underestimated error bars on some of the measurements.

To examine this issue we proceed as follows.  For each binning in the
weighted mean analysis above (i.e., with 9, 11, 13, and 16
measurements per bin) we compute the (``reduced $\chi^2$")
contribution to $\chi^2_B$ (eq. [4]) from each measurement in the bin,
$\chi^2_{B,i}$, where $\chi^2_B = \sum_{i=1}^{N_B} \chi^2_{B,i}$. We
then list the measurements in decreasing order of $\chi^2_{B,i}$ and
discard the first seven that appear in at least 3 of the 4 binnings
used. These ``discrepant" measurements are listed in Table 4. While
this procedure need not necessarily result in reducing all large
$N^B_\sigma$ values, it has the advantage of being less binning
dependent than a procedure designed solely to reduce large
$N^B_\sigma$ values to values that are consistent with the Gaussianity
assumption.

We first discuss the discrepant measurements of Table 4. As mentioned
above, the DMR results do not account for the full uncertainty in the
DMR data (G\'orski 1997; Gorski et al. 1998), and our analyses also
ignore correlations between the different DMR $\ell$-space
measurements. These effects might explain why the DMR measurements in
Table 4 are discrepant. The value of the cosmological DMR quadrupole
($\ell = 2$) moment is dependent on the model used to remove
foreground Galactic emission (Kogut et al. 1996), and this effect
might also contribute to explaining why the DMR $\ell = 2$ moment is
discrepant. The low MAX3 $\mu$ Pegasi result (Meinhold et al.  1993,
as recomputed by J. Gundersen, private communication 1995, see Ratra
et al. 1997) is from a region that is contaminated with dust, and this
effect could explain why this measurement is discrepant. While the
MAT97 Ka2 8-point and Q3 8-point results (Torbet et al. 1999) are
higher than neighboring measurements in $\ell$-space (see Table 1), we
do not know of an effect that might be responsible for making them
discrepant. The MSAM 2-beam combined result is from a combined
analysis of data from three different flights (Wilson et al. 2000). We
note that the three individual MSAM 2-beam results have significant
scatter (which seems to be larger than what one might expect from
their error bars, see Table 1 of Wilson et al. 2000, unlike the MSAM
3-beam results). It is unclear what causes this scatter, but it is
likely that this effect is responsible for placing the MSAM 2-beam
measurement among those that are discrepant. We again emphasize that
the measurements in Table 4 are discrepant only if the CMB anisotropy
is Gaussian. In particular, the posterior probability distribution
function of the DMR quadrupole is somewhat non-Gaussian (Hinshaw et
al. 1996), and this effect also contributes to explaining this
discrepancy.
     
After removing the seven most discrepant measurements we rebin the
remaining 135 measurements, using four different binnings, with about
9, 11, 13, and 15 measurements per $\ell$-space bin. We then analyze
this culled data set using both the weighted-mean and
median-statistics techniques. Tables 5 and 6 and Figures 4 and 5 show
the results from these analyses.

The weighted mean analyses (Figure 4 and Table 5) again result in
tight constraints on the observed CMB anisotropy power spectrum and
again clearly establish the presence of a peak.  This is again a
rather broad peak that lies in the interval $\ell \sim 170-240$ (for 9
measurements per bin, Figure 4$a$), $\ell \sim 170-260$ (for 11
measurements per bin, Figure 4$b$), $\ell \sim 200-250$ (for 13
measurements per bin, Figure 4$c$), and $\ell \sim 170-220$ (for 15
measurements per bin, Figure 4$d$). After removal of the discrepant
measurements of Table 4, the weighted-mean observed CMB anisotropy
power spectra have less scatter (compare Figures 4 and 2). There
remains the problem of 2 $\sigma$ upper limits that lie below the
observed spectrum of the detections, and in particular the $\ell =
138.7$ MAX5 $\mu$ Pegasi upper limit is now more inconsistent with the
observed spectrum of detections (see Figure 4).

While the culled data results in many fewer large $N^B_\sigma$ values
(the dashed line in each panel of Figure 4 and the last column in
Table 5), some bins still contain data that appear to be discrepant
(i.e., the weighted mean is more than 2 $\sigma$ away from what is
expected for a Gaussian distribution). $N^B_\sigma$ is larger than 2
for the penultimate of 15 bins for the 9 measurements per bin case,
for bin 6 and 11 (of 12) for the 11 measurements per bin case, for the
penultimate of 10 bins for the 13 measurements per bin case, and for
the penultimate of 9 bins for the 15 measurements per bin case. In the
penultimate bins, the most discrepant measurements are SK95 C15 and
MAT98 G9 (for 9 measurements per bin), SK95 C15 and MAXIMA-1 4 (for 11
measurements per bin), BOOMERanG98 8 and SK95 C13 (for 13 measurements
per bin), and BOOMERanG98 7, SK95 C12 and C13, the last two being
equally discrepant (for 15 measurements per bin). The two most
discrepant measurements in bin 6 for the 11 measurements per bin case
are MAX5 HR5127 6 cm$^{-1}$ and MAT97 Ka2 7-point. It would be useful
to understand why these measurements appear discrepant.

While it is possible to remove apparently discrepant measurements from
the already culled data set to reduce $N^B_\sigma$ below 2 for all
bins, it is not clear what is gained from this. Clearly, some
published CMB anisotropy measurements are mutually inconsistent, if
the CMB anisotropy is Gaussian and correlations between measurements
are not large. It would thus appear to be dangerous to draw
conclusions about the exact position of the peak in the CMB anisotropy
spectrum on the basis of the weighted mean analyses alone.
Furthermore, $\chi^2$ comparisons between CMB anisotropy measurements
and model predictions (to constrain cosmological parameter values) are
based on both the above assumptions. Such $\chi^2$ analyses of
collections of CMB anisotropy measurements most likely include
mutually inconsistent measurements, and so, without more careful
investigation, results based on such analyses must be viewed as
tentative, at best.

The median statistics results for the culled data set are shown in
Figure 5 (also see Table 6). Once again, the median statistics
analyses result in somewhat weaker constraints on the observed CMB
anisotropy detections power spectrum than do the weighted mean
analyses, but still clearly establish the presence of a peak. The
median-statistics peak interval is broader and typically extends more
to somewhat smaller angular scales than the weighted-mean peak
interval. The median-statistics peak lies in the interval $\ell \sim
150-280$ (for 9 measurements per bin, Figure 5$a$), $\ell
\sim 170-260$ (for 11 measurements per bin, Figure 5$b$), $\ell \sim 150-240$ 
(for 13 measurements per bin, Figure 5$c$), and $\ell \sim 170-300$ (for 15 
measurements per bin, Figure 5$d$). The median-statistics power spectra of 
CMB anisotropy detections are still inconsistent with some of the 2 $\sigma$ 
upper limits (see Figure 5).

Comparing the median-statistics observed CMB anisotropy detections
power spectra to model predictions (Figure 5), we see that the
$\Omega_0 = 0.4$ flat-$\Lambda$ and $\Omega_0 = 1$ fiducial CDM
inflation models are not inconsistent with the 2 $\sigma$ range of the
observations. While the fiducial CDM model might have slightly lower
power than is favored by the observations at $\ell \sim 200$ and at
$\ell \sim 400$, the $\Omega_0 = 0.4$ open CDM inflation model has
significantly lower power than is indicated by the observations at
$\ell \sim 100$ and at $\ell \sim 150-250$. We emphasize however that
the observations at $\ell \sim 300-500$ are quite consistent with the
(first peak of the) open model, and the error bars here are large
enough for the observations to be consistent with the presence of a
second peak in the flat-$\Lambda$ and fiducial CDM cases. Note that
results derived using the weighted mean technique (Figure 4) are
inconsistent with these predictions, i.e., at $\ell \sim 300-500$ the
weighted mean analyses rule out a first peak in the open case and a
second peak in the flat models.

\section{Conclusions}

We extend Page's (1997, 1999) weighted mean technique and use it to
determine the binned observed CMB anisotropy detection power
spectrum. Given the observational error bars, the binned power spectra
in a number of bins are not a good representation of the
measurements. Moreover, there are a number of 2 $\sigma$ upper limits
that are inconsistent with the binned power spectra of detections. A
number of effects could explain these results, but we suspect that a
major one is improperly accounted for or inadequately modelled
systematic effects, especially foreground emission contamination, with
correlations between some measurements playing a smaller role.

These results have two important ramifications. First, constraints on
cosmological parameters derived from $\chi^2$ comparisons between CMB
anisotropy model predictions and measurements must be interpreted with
great care. And certainly a full maximum likelihood comparison (e.g.,
Ratra et al. 1999b) is to be preferred. Second, it is important to
reanalyze observational data and account for systematic and other
effects that may have been ignored in the initial analyses (see Ganga
et al. 1997a for the general technique and Ratra et al. 1998, 1999a
for cases where a more careful accounting of such effects have led to
significant revisions of the observational results).

We have focussed on two methods for dealing with the inconsistencies
identified by the weighted mean analyses. We have used the median
statistics technique developed by Gott et al. (2001) to determine an
observed CMB anisotropy detections power spectrum. This method makes
fewer assumptions than the weighted mean technique (see discussion in
Gott et al. 2001). In particular, it completely ignores the error bars
on the individual measurements. The median statistics method results
in somewhat weaker constraints on the observed CMB anisotropy power
spectrum than does the weighted mean technique (compare Figures 3 and
2). If the CMB anisotropy is Gaussian, and if correlations between
measurements are small (as is expected for the data we have used),
then the weaker median-statistics constraints on the observed CMB
anisotropy power spectrum is another indication that the error bars on
some measurements might be too small (given the scatter in their
central values).

To test the impact of outliers, we remove the seven most discrepant
measurements (with largest reduced $\chi^2$ values, as discussed
above) and determine a weighted-mean observed CMB anisotropy power
spectrum (for the culled data set with 135 measurements). The
resultant power spectrum of detections is a much better fit for most
(but not all) bins but it is still inconsistent with some of the 2
$\sigma$ upper limits. A median statistics analysis of the culled data
set again results in significantly weaker constraints on the observed
power spectrum.

In summary, our analyses show that if the CMB anisotropy is Gaussian,
and if correlations between measurements are not significant, some
observational results are mutually inconsistent. Until this issue is
resolved it is not possible to determine a robust observational CMB
anisotropy power spectrum, much less constrain cosmological parameters
from $\chi^2$ comparisons of theoretical predictions and observational
results. While the open CDM model is not inconsistent with
observational data at $\ell \sim 300-500$, it predicts significantly
less power than is seen at $\ell \sim 150-250$. If foregrounds do not
contribute significantly to the data at $\ell \sim 150-250$, the
spatially-flat model must be favored significantly over the open
one. This result is consistent with other observational data that are
compatible with a flat model with a time-variable $\Lambda$ (see,
e.g., Ratra \& Peebles 1988; Haiman, Mohr, \& Holder 2000; Podariu \&
Ratra 2000; Gonz\'alez-D\'{\i}az 2000; Chimento, Jakubi, \& Pavo\'n
2000; Tye \& Wasserman 2001; Hebecker \& Wetterich 2001; Podariu,
Nugent, \& Ratra 2001; Barger \& Marfatia 2001). Unlike the
weighted-mean results, the median-statistics results do not rule out a
second peak in the CMB anisotropy at $\ell \sim 400-500$.

\bigskip

We acknowledge very valuable discussions with L. Page. We have
benefited from the advice and assistance of K. Coble, D. Cottingham,
P. de Bernardis, M. Devlin, S. Dicker, K. Ganga, K. G\'orski,
G. Griffin, J. Gundersen, S. Hanany, S. Hancock, E. Leitch, M. Lim,
P. Mukherjee, B. Netterfield, L. Piccirillo, S. Platt, D. Pogosyan,
G. Rocha, J. Ruhl, R. Stompor, N.  Sugiyama, S. Tanaka, E. Torbet, and
G. Tucker. SP, BR, and TS acknowledge support from NSF CAREER grant
AST-9875031. JRG acknowledges support from NSF grant AST-9900772. MSV
acknowledges support from the AAS Small Research Grant program, from
John Templeton Foundation grant 938-COS302, and from NSF grant
AST-0071201.

\clearpage

\begin{deluxetable}{lccccc}
\tablecaption{CMB Anisotropy Detections\tablenotemark{a}}
\tablewidth{0pt}
\tablehead{
\colhead{Experiment} &
\colhead{$\ell_{\rm e}$} &
\colhead{$\ell_{e^{-0.5}}$ range\tablenotemark{b}} &
\colhead{$\delta T_\ell$\tablenotemark{c}} &
\colhead{$\delta T_\ell$ (1 $\sigma$ range)\tablenotemark{d}} &
\colhead{Reference\tablenotemark{e}}
          }
\startdata
          &      &          &($\mu$K)& ($\mu$K)   & \\
\cline{1-6}
DMR       & 2.00 & 2.00--2.00 & 9.04 & 6.26--13.2 & G\'orski (1997) \\   
DMR       & 3.00 & 3.00--3.00 & 28.3 & 23.0--37.1 & \\
DMR       & 4.00 & 4.00--4.00 & 31.4 & 26.5--39.4 & \\
DMR       & 5.00 & 5.00--5.00 & 27.7 & 21.7--35.0 & \\
DMR       & 6.00 & 6.00--6.00 & 22.7 & 18.4--27.9 & \\  
DMR       & 7.00 & 7.00--7.00 & 22.1 & 16.0--29.0 & \\ 
DMR       & 8.00 & 8.00--8.00 & 23.5 & 18.7--28.9 & \\
DMR       & 9.00 & 9.00--9.00 & 40.0 & 34.1--47.0 & \\ 
DMR       & 10.0 & 10.0--10.0 & 26.2 & 21.1--32.1 & \\  
FIRS      & 10.8 &[2.00]--25.0& 31.4 & 23.4--39.3 & Bond (1995) \\
DMR       & 11.0 & 11.0--11.0 & 40.2 & 34.0--47.5 & \\
DMR       & 12.0 & 12.0--12.0 & 20.4 & 13.1--27.4 & \\  
DMR       & 13.0 & 13.0--13.0 & 40.1 & 33.7--47.2 & \\ 
DMR       & 14.0 & 14.0--14.0 & 32.1 & 25.8--38.8 & \\
DMR       & 17.0 & 17.0--17.0 & 55.1 & 47.0--64.0 & \\  
DMR       & 18.0 & 18.0--18.0 & 38.3 & 28.4--48.6 & \\  
Tenerife  & 20.1 & 14.0--29.0 & 30.0 & 18.6--45.3 & Guti\'errez et al. (2000) \\
PyVM1     & 49.6 & 28.0--60.0 & 27.7 & 22.3--34.0 & T. Souradeep pvt. comm. (2000)\tablenotemark{f} \\
BOOM98-1  & 50.0 & 26.0--75.0 & 33.8 & 28.2--39.0 & de Bernardis et al. (2000) \\
IACBartol & 52.7 & 38.0--77.0 & 54.6 & 32.7--81.8 & Femen\'{\i}a et al. (1998) \\
SK94Ka3   & 55.8 & 37.0--76.0 & 51.4 & 41.2--70.3 & Netterfield et al. (1997) \\
SP94Ka    & 57.2 & 35.0--98.0 & 30.6 & 21.8--43.5 & Ganga et al. (1997a) \\
BOOM97-1  & 58.0 & 25.0--75.0 & 29.2 & 17.4--42.1 & Mauskopf et al. (2000) \\
BAM       & 58.2 & 16.0--92.0 & 55.6 & 40.8--85.2 & Tucker et al. (1997) \\
SK94Q3    & 59.4 & 37.0--76.0 & 41.8 & 28.4--63.1 & \\
MAT97Ka2-4& 63.4 & 45.0--82.0 & 35.0 & 25.3--48.5 & Torbet et al. (1999) \\
MAT97Q1-4 & 63.4 & 46.0--82.0 & 57.0 & 42.8--75.9 & \\
SK95C3    & 63.7 & 37.0--76.0 & 67.0 & 50.8--92.7 & Netterfield et al. (1997) \\
SP94Q     & 66.2 & 40.0--112. & 38.8 & 28.9--52.7 & \\
SK93      & 67.0 & 47.0--95.0 & 37.1 & 27.0--50.1 & Netterfield et al. (1997) \\
PyVM2     & 73.5 & 48.0--100. & 30.8 & 24.2--38.3 & \\
SK94Ka4   & 76.5 & 57.0--96.0 & 33.1 & 25.5--46.8 & \\
MAXIMA1-1 & 77.0 & 36.0--110. & 44.7 & 38.3--52.0 & Hanany et al. (2000) \\
MAT97Q3-5 & 82.7 & 63.0--100. & 47.0 & 33.2--64.6 & \\
SK95C4    & 83.0 & 54.0--96.0 & 39.1 & 29.8--56.2 & \\
QMAP1Q    & 83.7 & 43.0--146. & 47.0 & 35.5--56.8 & Devlin et al. (1998) \\
MAT97Ka2-5& 86.5 & 65.0--103. & 52.0 & 42.5--64.2 & \\
MAT97Q1-5 & 86.7 & 65.0--104. & 40.0 & 25.4--54.6 & \\
PyIIIL    & 87.7 & 52.0--98.0 & 63.7 & 49.4--84.5 & Rocha et al. (1999) \\
QMAP2Ka1  & 91.1 & 60.0--165. & 44.9 & 32.0--56.1 & Herbig et al. (1998) \\
PyI+II    & 91.7 & 53.0--99.0 & 52.1 & 39.0--71.6 & Rocha et al. (1999) \\
QMAP1Ka   & 92.0 & 56.0--159. & 47.9 & 39.2--55.8 & \\
SK94Ka5   & 96.0 & 76.0--116. & 45.1 & 34.6--63.4 & \\
ARGO Herc.& 97.6 & 60.0--168. & 32.6 & 28.0--37.9 & Ratra et al. (1999a) \\
BOOM98-2  & 100. & 76.0--125. & 55.8 & 48.5--62.8 & \\
BOOM97-2  & 102. & 76.0--125. & 48.8 & 39.2--58.9 & \\
MAT97Q3-6 & 106. & 84.0--121. & 61.0 & 46.6--80.0 & \\
JB-IAC-L  & 106. & 90.0--129. & 43.0 & 30.7--55.8 & Dicker et al. (1999) \\
MAT97Ka2-6& 107. & 86.0--124. & 71.0 & 58.7--84.9 & \\
SK95C5    & 108. & 76.0--115. & 55.6 & 43.6--76.1 & \\
PyVM3     & 108. & 81.0--131. & 33.5 & 26.2--41.7 & \\
Viper1    & 108. & 46.0--135. & 61.0 & 38.5--92.4 & Peterson et al. (2000) \\
MAT97Q1-6 & 110. & 87.0--126. & 56.0 & 41.9--71.1 & \\
MAX4 6/cm $\gamma$UM &114.&70.0--196.&41.6&29.2--60.3&S. Tanaka pvt. comm. (1995)\\
MAX4 6/cm $\iota$D   &114.&70.0--196.&67.5&37.6--112.&Ganga et al. (1998)\\
MAX4 9/cm $\gamma$UM &114.&70.0--196.&53.6&37.0--79.2&S. Tanaka pvt. comm. (1995)\\
MAX4 9/cm $\sigma$H  &114.&70.0--196.&53.0&29.6--90.7&Ganga et al. (1998)\\   
SK94Ka6   & 115. & 95.0--136. & 34.3 & 22.8--50.9 & \\
QMAP2Q    & 125. & 76.0--228. & 56.0 & 47.4--63.9 & \\
MAT97Q3-7 & 125. & 103.--141. & 72.0 & 58.0--89.6 & \\
MAT97Ka2-7& 127. & 106.--145. & 93.0 & 76.2--111. & \\
MAT98G6   & 129. & 96.0--153. & 55.0 & 37.4--73.5 & Miller et al. (1999) \\
MAT97Q1-7 & 131. & 107.--147. & 81.0 & 63.1--102. & \\
MAX4 3.5/cm $\gamma$UM&133.&80.0--224.&78.6&56.3--108.&S. Tanaka pvt. comm. (1995)\\
MAX4 3.5/cm $\sigma$H &133.&80.0--224.&85.8&58.4--129.&Ganga et al. (1998)\\  
MAX4 3.5/cm $\iota$D  &133.&80.0--224.&56.6&37.4--86.4&Ganga et al. (1998)\\ 
MAX5 6/cm HR          &133.&80.0--224.&26.7&19.7--37.1&Ganga et al. (1998)\\  
MAX5 6/cm $\phi$H     &133.&80.0--224.&73.8&55.9--99.4&Ganga et al. (1998)\\  
MAX5 9/cm HR          &133.&80.0--224.&37.0&24.6--54.6&Ganga et al. (1998)\\  
SK95C6    & 135. & 97.0--141. & 65.9 & 52.5--85.8 &  \\
MAX5 3.5/cm HR        &139.&83.0--232.&40.0&27.5--58.0&Ganga et al. (1998)\\    
MAX5 3.5/cm $\phi$H   &139.&83.0--232.&50.5&32.0--77.2&Ganga et al. (1998)\\    
PyVM4     & 141. & 113.--161. & 37.8 & 26.8--49.3 &  \\
MAX3 $\gamma$UM       &142.&85.0--240.&74.2&59.8--96.1&J. Gundersen pvt. comm. (1995)\\
MAX3 $\mu$Peg         &144.&86.0--243.&23.4&16.2--36.0&J. Gundersen pvt. comm. (1995)\\
MAT97Ka2-8& 145. & 126.--165. & 103. & 86.4--121. & \\
QMAP2Ka2  & 145. & 141.--224. & 62.2 & 48.2--74.6 & \\
MAT97Q3-8 & 145. & 122.--161. & 115. & 96.1--137. & \\
MAXIMA1-2 & 147. & 111.--185. & 54.4 & 48.7--60.7 & \\
BOOM98-3  & 150. & 126.--175. & 64.5 & 56.5--72.2 & \\
MAT97Q1-8 & 151. & 127.--167. & 86.0 & 67.0--107. & \\ 
BOOM97-3  & 153. & 126.--175. & 67.2 & 56.6--78.2 & \\
MAT98G7   & 155. & 117.--183. & 82.0 & 69.2--94.8 & \\
SK95C7    & 158. & 119.--165. & 74.2 & 59.4--95.4 & \\
MSAM 2beam comb.     &159.&83.0--234.&47.0&41.5--52.5& Wilson et al. (2000) \\
MAT97Q3-9 & 165. & 141.--180. & 72.0 & 49.8--97.1 & \\
MAT97Ka2-9& 166. & 145.--184. & 65.0 & 46.8--82.3 & \\
SK95R3    & 170. & 115.--236. & 60.8 & 49.3--73.9 & \\
PyIIIS    & 171. & 128.--230. & 65.7 & 51.4--86.6 & Rocha et al. (1999) \\
PyVM5     & 172. & 145.--192. & 58.4 & 41.4--75.9 & \\
MAT97Q1-9 & 172. & 147.--188. & 93.0 & 68.2--118. & \\
Viper2    & 173. & 92.0--193. & 77.0 & 56.1--104. & \\
SK94Q9    & 176. & 153.--195. & 142. & 94.5--204. & \\
SK95C8    & 178. & 140.--184. & 83.4 & 68.0--106. & \\
MAT97Ka2-10&182. & 165.--204. & 67.0 & 43.0--88.1 & \\ 
MAT97Q3-10& 184. & 161.--200. & 87.0 & 66.1--108. & \\ 
MAT97Q3-11& 196. & 180.--219. & 90.0 & 62.5--119. & \\ 
SK95C9    & 197. & 159.--204. & 78.3 & 62.4--101. & \\ 
BOOM98-4  & 200. & 176.--225. & 68.6 & 60.3--76.7 & \\ 
PyVM6     & 203. & 176.--223. & 95.4 & 68.8--123. & \\ 
BOOM97-4  & 204. & 176.--225. & 71.9 & 60.5--83.3 & \\ 
JB-IAC-S  & 207. & 190.--227. & 63.0 & 55.7--71.1 & Harrison et al. (2000) \\
MAT97Q3-12& 212. & 199.--238. & 100. & 71.2--131. & \\
MAT97Ka2-12&215. & 204.--243. & 128. & 92.6--161. & \\ 
SK95C10   & 217. & 178.--222. & 78.3 & 59.7--103. & \\
MAXIMA1-3 & 223. & 186.--260. & 77.9 & 71.1--85.1 & \\
SK95R4    & 234. & 182.--301. & 80.3 & 65.2--99.  & \\
SK95C11   & 237. & 199.--244. & 85.5 & 67.4--110. & \\ 
Viper3    & 237. & 148.--283. & 65.0 & 47.2--89.6 & \\ 
MAT98G8   & 248. & 185.--302. & 83.0 & 72.6--92.7 & \\ 
BOOM98-5  & 250. & 226.--275. & 65.6 & 57.7--73.4 & \\
BOOM97-5  & 255. & 226.--275. & 60.8 & 48.0--73.1 & \\ 
SK95C12   & 257. & 221.--265. & 115. & 92.9--146. & \\ 
MAT97Q3-14& 258. & 236.--277. & 119. & 79.2--157. & \\ 
Viper4    & 263. & 157.--441. & 79.0 & 63.6--98.1 & \\ 
MSAM 3beam comb.     &263.&181.--375.&53.0&47.3--58.7& Wilson et al. (2000) \\  
SK95C13   & 277. & 241.--286. & 119. & 94.9--151. & \\ 
SK95R5    & 286. & 247.--365. & 71.1 & 48.1--93.1 & \\
SK95C14   & 297. & 263.--307. & 76.2 & 50.1--104. & \\ 
MAXIMA1-4 & 300. & 261.--335. & 61.0 & 55.7--66.5 & \\ 
BOOM98-6  & 300. & 276.--325. & 51.4 & 45.0--57.7 & \\ 
BOOM97-6  & 305. & 276.--325. & 55.4 & 38.6--69.7 & \\ 
SK95C15   & 316. & 282.--326. & 128. & 97.4--166. & \\ 
MAT98G9   & 319. & 267.--347. & 70.0 & 57.7--81.5 & \\
SK95C16   & 334. & 301.--345. & 113. & 72.2--154. & \\
BOOM98-7  & 350. & 326.--375. & 39.4 & 34.0--44.7 & \\
MAXIMA1-5 & 374. & 336.--410. & 47.6 & 43.4--52.1 & \\
CAT1 (Yr. 1)&396.& 351.--471. & 51.8 & 37.9--65.7 & G. Rocha pvt. comm. (1997)\\
CAT1 (Yr. 2)&396.& 351.--471. & 56.6 & 42.6--69.6 & Baker et al. (1999) \\
BOOM98-8  & 400. & 376.--425. & 36.2 & 30.7--41.4 & \\
MAXIMA1-6 & 447. & 411.--485. & 39.1 & 35.1--43.3 & \\
BOOM98-9  & 450. & 426.--475. & 36.8 & 30.8--42.5 & \\
BOOM98-10 & 500. & 476.--525. & 38.0 & 31.2--44.2 & \\
MAXIMA1-7 & 522. & 486.--560. & 48.4 & 43.6--53.3 & \\
BOOM98-11 & 550. & 526.--575. & 41.8 & 33.9--49.0 & \\
Viper6    & 588. & 443.--794. & 65.0 & 39.5--90.5 & \\
MAXIMA1-8 & 597. & 561.--635. & 39.1 & 34.1--44.2 & \\
OVRO      & 599. & 361.--754. & 59.0 & 52.5--67.6 & Leitch et al. (2000) \\
BOOM98-12 & 600. & 576.--625. & 39.2 & 30.0--47.2 & \\
CAT2 (Yr. 1)&608.& 565.--710. & 49.1 & 35.2--68.3 & G. Rocha pvt. comm. (1997)\\
MAXIMA1-9 & 671. & 636.--710. & 42.8 & 36.7--48.8 & \\
MAXIMA1-10& 742. & 711.--785. & 46.7 & 38.9--54.4 & 
\enddata
\tablenotetext{a}{Where known, beamwidth and calibration uncertainties have
  been accounted for and foreground contamination removed.} 
\tablenotetext{b}{The two values of $\ell$ where $W_{\ell_{e^{-0.5}}} =
  e^{-0.5} W_{\ell_{\rm m}}$, where $\ell_{\rm m}$ is the value of $\ell$ 
  where $W_\ell$ is largest.}
\tablenotetext{c}{Bandtemperature central value (this is where the likelihood
  is largest), typically derived assuming a flat bandpower spectrum.}
\tablenotetext{d}{1 $\sigma$ range of band temperature. Accounts for known 
  systematic uncertainties.}
\tablenotetext{e}{Typically provided only the first time the experiment
  appears in the Table.} 
\tablenotetext{f}{Ignores cross-modulation correlations.}
\end{deluxetable}

\begin{deluxetable}{lcccccc}
\tablecaption{Weighted Mean Results Using All Measurements}
\tablewidth{0pt}
\tablehead{
\colhead{Bin} &
\colhead{$\ell_{\rm e}^B${\,}\tablenotemark{a}} &
\colhead{$N^B$} &
\colhead{$\delta T_\ell^B$} &
\colhead{$\delta T_\ell^B$ (1 $\sigma$ range)} &
\colhead{$\delta T_\ell^B$ (2 $\sigma$ range)} &
\colhead{$N_\sigma^B$}
}
\startdata
 & & & ($\mu$K) & ($\mu$K) & ($\mu$K) & \\
\cutinhead{9 measurements per bin}
1  & 4.97 &  8 & 21.8 & 19.9--23.7 & 18.0--25.6 & 3.39 \\
2  & 12.8 &  9 & 33.8 & 31.4--36.3 & 28.9--38.7 & 1.29 \\
3  & 52.0 &  8 & 33.3 & 29.9--36.6 & 26.6--39.9 & .691 \\
4  & 71.8 &  8 & 38.9 & 35.3--42.4 & 31.7--46.0 & .432 \\ 
5  & 87.8 &  9 & 47.6 & 43.5--51.7 & 39.5--55.7 & 2.09 \\
6  & 102. & 10 & 43.3 & 40.3--46.3 & 37.4--49.3 & 1.46 \\
7  & 119. &  8 & 53.1 & 47.9--58.4 & 42.7--63.6 & .885 \\
8  & 132. & 10 & 51.9 & 46.6--57.1 & 41.4--62.3 & 2.26 \\
9  & 145. &  9 & 52.8 & 48.9--56.7 & 45.0--60.6 & 3.83 \\
10 & 156. &  8 & 59.8 & 56.2--63.5 & 52.5--67.2 & 1.14 \\
11 & 174. &  9 & 71.3 & 64.8--77.7 & 58.3--84.2 & .882 \\ 
12 & 204. &  9 & 71.2 & 66.7--75.8 & 62.2--80.3 & .408 \\
13 & 240. &  9 & 75.0 & 71.1--79.0 & 67.1--83.0 & .0803\\
14 & 287. &  8 & 57.7 & 54.6--60.9 & 51.5--64.0 & .529 \\
15 & 402. & 10 & 42.5 & 40.5--44.6 & 38.4--46.7 & 2.19 \\
16 & 588. & 10 & 44.4 & 42.2--46.6 & 39.9--48.8 & .285 \\
\cutinhead{11 measurements per bin}
1  & 6.09 & 11 & 23.8 & 22.1--25.5 & 20.4--27.2 & 3.72 \\
2  & 29.8 & 10 & 34.4 & 31.9--36.8 & 29.4--39.3 & 1.10 \\
3  & 68.9 & 12 & 37.9 & 34.8--41.1 & 31.6--44.3 & .897 \\
4  & 88.4 & 10 & 47.4 & 43.5--51.3 & 39.6--55.2 & 2.32 \\
5  & 102. & 10 & 43.8 & 40.8--46.8 & 37.8--49.7 & 1.58 \\
6  & 123. & 10 & 58.0 & 53.1--63.0 & 48.2--67.9 & .337 \\
7  & 137. & 12 & 40.7 & 36.4--45.0 & 32.1--49.2 & 1.33 \\
8  & 153. & 12 & 61.0 & 58.1--64.0 & 55.1--67.0 & 2.51 \\
9  & 177. & 11 & 72.8 & 66.8--78.8 & 60.8--84.8 & 1.19 \\
10 & 213. & 11 & 73.2 & 69.5--76.9 & 65.8--80.6 & .645 \\
11 & 277. & 12 & 61.7 & 59.0--64.5 & 56.2--67.2 & 1.88 \\
12 & 400. & 11 & 42.8 & 40.7--44.8 & 38.6--46.9 & 2.07 \\
13 & 588. & 10 & 44.4 & 42.2--46.6 & 39.9--48.8 & .285 \\
\cutinhead{13 measurements per bin}
1  & 6.09 & 11 & 23.8 & 22.1--25.5 & 20.4--27.2 & 3.72 \\
2  & 32.4 & 13 & 34.2 & 31.9--36.6 & 29.6--38.9 & .670 \\
3  & 73.5 & 12 & 40.0 & 36.9--43.2 & 33.7--46.3 & 1.16 \\
4  & 96.3 & 12 & 44.7 & 41.8--47.5 & 39.0--50.3 & .123 \\
5  & 116. & 14 & 52.2 & 48.4--56.0 & 44.7--59.7 & .870 \\
6  & 136. & 12 & 46.7 & 42.1--51.3 & 37.5--55.9 & 1.28 \\
7  & 152. & 13 & 58.0 & 55.1--60.8 & 52.3--63.7 & 3.94 \\
8  & 188. & 14 & 72.0 & 67.7--76.4 & 63.3--80.8 & 1.55 \\
9  & 241. & 15 & 68.3 & 65.4--71.2 & 62.5--74.1 & 1.51 \\
10 & 349. & 14 & 49.6 & 47.4--51.7 & 45.2--53.9 & 3.09 \\
11 & 546. & 12 & 42.6 & 40.7--44.4 & 38.9--46.3 & .181 \\
\cutinhead{16 measurements per bin}
1 & 7.89 & 17 & 26.3 & 24.8--27.8 & 23.3--29.3 & 4.79 \\
2 & 57.8 & 14 & 34.7 & 32.0--37.3 & 29.3--40.0 & .742 \\
3 & 90.7 & 15 & 43.7 & 41.3--46.2 & 38.8--48.7 & .413 \\
4 & 113. & 14 & 49.9 & 46.2--53.6 & 42.5--57.2 & .0767\\
5 & 138. & 18 & 51.8 & 48.1--55.4 & 44.5--59.1 & 4.33 \\
6 & 156. & 14 & 59.5 & 56.5--62.4 & 53.6--65.3 & .401 \\
7 & 210. & 17 & 74.4 & 71.0--77.9 & 67.6--81.3 & 1.19 \\
8 & 294. & 17 & 58.0 & 55.7--60.4 & 53.3--62.7 & 3.57 \\
9 & 505. & 16 & 43.0 & 41.4--44.6 & 39.8--46.2 & .111 
\enddata
\tablenotetext{a}{Weighted mean of $\ell_{\rm e}$ values of measurements in 
  the bin.} 
\end{deluxetable}

\begin{deluxetable}{lccccc}
\tablecaption{Median Statistics Results Using All Measurements}
\tablewidth{0pt}
\tablehead{
\colhead{Bin} &
\colhead{$\ell_{\rm e}^B${\,}\tablenotemark{a}} &
\colhead{$N^B$} &
\colhead{$\delta T_\ell^B$} &
\colhead{$\delta T_\ell^B$ (1 $\sigma$ range)} &
\colhead{$\delta T_\ell^B$ (2 $\sigma$ range)} 
}
\startdata
 & & & ($\mu$K) & ($\mu$K) & ($\mu$K) \\
\cutinhead{9 measurements per bin\tablenotemark{b}}
1  & 5.50 &  8 & 24.7 & 22.6--28.1 & 11.4--31.9 \\
2  & 13.0 &  9 & 31.9 & 30.3--39.0 & 24.2--40.2 \\
3  & 56.5 &  8 & 35.2 & 30.2--48.1 & 28.1--54.6 \\
4  & 66.6 &  8 & 37.6 & 34.8--43.1 & 31.5--56.8 \\
5  & 86.7 &  9 & 47.0 & 45.6--49.8 & 39.9--52.1 \\
6  & 106. & 10 & 50.7 & 44.3--55.8 & 34.2--61.0 \\
7  & 114. &  8 & 54.5 & 50.7--56.0 & 35.8--68.1 \\
8  & 133. & 10 & 67.6 & 55.8--78.1 & 35.6--83.7 \\
9  & 144. &  9 & 53.0 & 41.0--64.3 & 29.3--93.2 \\
10 & 156. &  8 & 68.9 & 64.9--73.8 & 50.4--82.8 \\
11 & 173. &  9 & 74.8 & 66.0--84.8 & 60.1--91.9 \\
12 & 204. &  9 & 78.3 & 73.6--92.1 & 67.0--99.3 \\
13 & 248. &  9 & 79.9 & 68.5--84.0 & 63.8--109. \\
14 & 292. &  8 & 63.6 & 55.0--75.0 & 51.8--79.0 \\
15 & 385. & 10 & 48.4 & 39.3--55.5 & 37.1--78.3 \\
16 & 598. & 10 & 44.2 & 41.1--48.3 & 39.1--54.6 \\
\cutinhead{11 measurements per bin\tablenotemark{b}}
1  & 7.00 & 11 & 27.4 & 23.8--29.9 & 22.3--32.0 \\
2  & 19.1 & 10 & 34.9 & 31.2--39.9 & 27.8--52.9 \\
3  & 63.6 & 12 & 37.6 & 34.0--41.9 & 30.7--52.8 \\
4  & 87.2 & 10 & 47.0 & 45.1--47.9 & 40.7--52.1 \\
5  & 107. & 10 & 55.7 & 46.2--56.0 & 33.7--61.0 \\
6  & 120. & 10 & 55.3 & 53.4--66.0 & 42.0--76.6 \\
7  & 134. & 12 & 50.9 & 38.5--64.7 & 32.2--74.1 \\
8  & 152. & 12 & 68.8 & 64.8--74.8 & 61.3--85.2 \\
9  & 176. & 11 & 78.1 & 68.2--85.5 & 62.7--90.5 \\
10 & 215. & 11 & 78.2 & 72.8--83.3 & 66.5--96.0 \\
11 & 263. & 12 & 72.5 & 63.2--79.1 & 59.5--106. \\
12 & 374. & 11 & 50.6 & 39.8--56.0 & 37.6--68.8 \\
13 & 598. & 10 & 44.2 & 41.1--48.3 & 39.1--54.6 \\
\cutinhead{13 measurements per bin\tablenotemark{b}}
1  & 7.00 & 11 & 27.4 & 23.8--29.9 & 22.3--32.0 \\
2  & 49.6 & 13 & 33.3 & 30.6--39.3 & 29.3--52.4 \\
3  & 70.3 & 12 & 40.1 & 38.0--44.9 & 34.8--47.0 \\
4  & 94.0 & 12 & 48.2 & 45.0--52.0 & 42.8--55.2 \\
5  & 114. & 14 & 55.8 & 54.3--57.4 & 46.8--67.5 \\
6  & 133. & 12 & 57.6 & 42.8--73.3 & 37.5--77.6 \\
7  & 151. & 13 & 66.8 & 64.5--73.5 & 55.4--83.9 \\
8  & 180. & 14 & 77.4 & 69.9--84.4 & 66.4--90.2 \\
9  & 237. & 15 & 78.9 & 76.0--82.5 & 65.0--95.6 \\
10 & 317. & 14 & 57.7 & 53.1--70.1 & 49.4--75.7 \\
11 & 592. & 12 & 42.2 & 39.2--46.9 & 39.1--49.0 \\
\cutinhead{16 measurements per bin\tablenotemark{b}}
1 & 10.0 & 17 & 29.7 & 27.2--31.4 & 23.2--37.6 \\
2 & 58.8 & 14 & 37.5 & 34.2--42.5 & 30.7--54.4 \\
3 & 87.7 & 15 & 46.8 & 44.9--47.8 & 40.2--51.4 \\
4 & 112. & 14 & 55.8 & 53.3--57.5 & 42.2--61.5 \\
5 & 134. & 18 & 62.8 & 55.0--73.9 & 39.6--79.9 \\ 
6 & 162. & 14 & 66.3 & 64.7--72.6 & 59.8--77.3 \\
7 & 204. & 17 & 80.1 & 78.2--85.4 & 71.0--89.9 \\
8 & 286. & 17 & 70.8 & 64.0--78.5 & 58.7--106. \\ 
9 & 536. & 16 & 44.0 & 40.0--47.9 & 39.1--50.2
\enddata
\tablenotetext{a}{Median of $\ell_{\rm e}$ values of measurements in the bin.}
\tablenotetext{b}{The lower cutoffs on the integral of the likelihood function
  are 4, 1, 1, and 1 $\mu$K for 9, 11, 13, and 16 measurements per bin.}
\end{deluxetable}

\begin{deluxetable}{lccccc}
\tablecaption{Discrepant Measurements}
\tablewidth{0pt}
\tablehead{
\colhead{Measurement} &
\colhead{$\ell$} &
\colhead{$\chi^2_{B,i}$(9)\tablenotemark{a}} &
\colhead{$\chi^2_{B,i}$(11)\tablenotemark{a}} &
\colhead{$\chi^2_{B,i}$(13)\tablenotemark{a}} &
\colhead{$\chi^2_{B,i}$(16)\tablenotemark{a}}
}
\startdata
DMR               & 2.00 & 1.9 & 1.8 & 1.8 & 1.5 \\
DMR               & 9.00 & 1.1 & .64 & .64 & .29 \\
DMR               & 17.0 & .79 & .66 & .50 & .72 \\
MAX3 $\mu$Peg     & 144. & 1.1 & .27 & 1.0 & .48 \\
MAT97Ka2-8        & 145. & 1.0 & .53 & .56 & .51 \\
MAT97Q3-8         & 145. & 1.1 & .63 & .64 & .56 \\
MSAM 2beam comb.  & 159. & .77 & .59 & .33 & .39
\enddata
\tablenotetext{a}{Reduced $\chi^2$ of the measurement for the case with
  9, 11, 13, and 16 measurements per bin, respectively.}
\end{deluxetable}

\begin{deluxetable}{lcccccc}
\tablecaption{Weighted Mean Results Using ``Good" Measurements Only}
\tablewidth{0pt}
\tablehead{
\colhead{Bin} &
\colhead{$\ell_{\rm e}^B${\,}\tablenotemark{a}} &
\colhead{$N^B$} &
\colhead{$\delta T_\ell^B$} &
\colhead{$\delta T_\ell^B$ (1 $\sigma$ range)} &
\colhead{$\delta T_\ell^B$ (2 $\sigma$ range)} &
\colhead{$N_\sigma^B$}
}
\startdata
 & & & ($\mu$K) & ($\mu$K) & ($\mu$K) & \\
\cutinhead{9 measurements per bin}
1  & 7.18 &  9 & 27.2 & 25.1--29.2 & 23.1--31.2 & .394 \\
2  & 30.2 &  8 & 31.8 & 29.2--34.5 & 26.6--37.1 & .275 \\
3  & 59.6 &  8 & 40.7 & 35.6--45.8 & 30.5--50.9 & .417 \\
4  & 75.7 &  8 & 38.9 & 35.5--42.4 & 32.0--45.8 & 1.20 \\
5  & 94.9 &  9 & 43.8 & 40.7--46.9 & 37.7--49.9 & .550 \\
6  & 107. &  8 & 48.0 & 43.6--52.3 & 39.3--56.7 & .293 \\
7  & 122. &  9 & 56.4 & 51.3--61.5 & 46.2--66.6 & .237 \\
8  & 134. & 10 & 46.3 & 41.0--51.5 & 35.8--56.8 & 1.21 \\
9  & 148. &  9 & 61.3 & 57.8--64.8 & 54.2--68.3 & .749 \\
10 & 171. &  9 & 69.4 & 63.0--75.8 & 56.6--82.1 & 1.07 \\
11 & 202. &  9 & 70.5 & 66.0--75.0 & 61.6--79.4 & 1.13 \\
12 & 238. &  9 & 74.5 & 70.6--78.5 & 66.6--82.4 & .388 \\
13 & 287. & 10 & 58.9 & 55.8--62.0 & 52.7--65.1 & 1.45 \\
14 & 402. & 10 & 42.5 & 40.5--44.6 & 38.4--46.7 & 2.19 \\
15 & 588. & 10 & 44.4 & 42.2--46.6 & 39.9--48.8 & .285 \\
\cutinhead{11 measurements per bin}
1  & 7.96 & 11 & 27.7 & 25.8--29.6 & 24.0--31.4 & .225 \\
2  & 41.1 & 11 & 33.3 & 30.5--36.1 & 27.7--38.8 & 1.31 \\
3  & 74.0 & 11 & 40.0 & 36.8--43.2 & 33.5--46.4 & .947 \\
4  & 96.3 & 12 & 44.7 & 41.8--47.5 & 39.0--50.3 & .123 \\
5  & 110. & 10 & 46.1 & 41.4--50.7 & 36.7--55.4 & .0751\\
6  & 130. & 12 & 54.4 & 50.2--58.7 & 45.9--63.0 & 2.04 \\
7  & 148. & 11 & 60.0 & 56.6--63.4 & 53.2--66.8 & .688 \\
8  & 172. & 11 & 70.6 & 64.7--76.5 & 58.9--82.4 & 1.32 \\
9  & 211. & 11 & 73.6 & 69.9--77.3 & 66.2--81.0 & .661 \\
10 & 257. & 11 & 65.9 & 62.3--69.5 & 58.7--73.1 & 1.63 \\
11 & 350. & 12 & 48.9 & 46.7--51.1 & 44.5--53.3 & 2.59 \\
12 & 546. & 12 & 42.6 & 40.7--44.4 & 38.9--46.3 & .181 \\
\cutinhead{13 measurements per bin}
1  & 8.92 & 14 & 28.4 & 26.6--30.1 & 24.9--31.9 & .104 \\
2  & 55.1 & 13 & 35.3 & 32.4--38.2 & 29.5--41.1 & .624 \\
3  & 86.6 & 14 & 40.2 & 37.7--42.8 & 35.2--45.3 & .563 \\
4  & 106. & 14 & 49.0 & 45.6--52.4 & 42.2--55.9 & .469 \\
5  & 131. & 14 & 53.3 & 49.3--57.3 & 45.2--61.4 & 1.75 \\
6  & 152. & 13 & 61.7 & 58.5--65.0 & 55.2--68.2 & .0879\\
7  & 197. & 13 & 71.1 & 67.0--75.2 & 62.8--79.3 & 1.25 \\
8  & 247. & 14 & 69.2 & 66.1--72.4 & 63.0--75.5 & 1.62 \\
9  & 349. & 14 & 49.6 & 47.4--51.7 & 45.2--53.9 & 3.09 \\
10 & 546. & 12 & 42.6 & 40.7--44.4 & 38.9--46.3 & .181 \\
\cutinhead{15 measurements per bin}
1 & 8.92 & 14 & 28.4 & 26.6--30.1 & 24.9--31.9 & .104 \\
2 & 61.2 & 16 & 35.9 & 33.4--38.3 & 31.0--40.7 & .693 \\
3 & 94.7 & 15 & 44.6 & 42.0--47.3 & 39.4--49.9 & .427 \\
4 & 116. & 15 & 53.3 & 49.6--56.9 & 45.9--60.6 & 1.03 \\
5 & 142. & 15 & 52.9 & 49.6--56.1 & 46.4--59.3 & 1.49 \\
6 & 166. & 14 & 71.7 & 67.1--76.3 & 62.4--81.0 & 1.68 \\
7 & 224. & 16 & 72.6 & 69.6--75.6 & 66.5--78.6 & .879 \\
8 & 322. & 15 & 53.4 & 51.1--55.6 & 48.9--57.8 & 3.27 \\
9 & 526. & 15 & 42.3 & 40.5--44.0 & 38.8--45.7 & .177
\enddata
\tablenotetext{a}{Weighted mean of $\ell_{\rm e}$ values of measurements in 
  the bin.} 
\end{deluxetable}

\begin{deluxetable}{lccccc}
\tablecaption{Median Statistics Results Using ``Good" Measurements Only}
\tablewidth{0pt}
\tablehead{
\colhead{Bin} &
\colhead{$\ell_{\rm e}^B${\,}\tablenotemark{a}} &
\colhead{$N^B$} &
\colhead{$\delta T_\ell^B$} &
\colhead{$\delta T_\ell^B$ (1 $\sigma$ range)} &
\colhead{$\delta T_\ell^B$ (2 $\sigma$ range)} 
}
\startdata
 & & & ($\mu$K) & ($\mu$K) & ($\mu$K) \\
\cutinhead{9 measurements per bin\tablenotemark{b}}
1  & 7.00 &  9 & 27.4 & 24.2--29.6 & 22.6--31.4 \\
2  & 19.1 &  8 & 32.6 & 29.6--37.3 & 22.0--40.8 \\
3  & 58.8 &  8 & 43.4 & 33.5--54.4 & 29.4--57.6 \\
4  & 76.7 &  8 & 38.9 & 36.6--43.7 & 31.4--47.0 \\
5  & 91.7 &  9 & 47.5 & 45.0--52.0 & 37.5--55.3 \\
6  & 108. &  8 & 55.7 & 47.4--60.1 & 35.0--62.9 \\
7  & 115. &  9 & 54.8 & 53.2--60.2 & 38.4--71.2 \\
8  & 133. & 10 & 57.9 & 44.4--72.4 & 37.0--79.8 \\
9  & 150. &  9 & 66.7 & 62.7--74.2 & 46.0--81.0 \\
10 & 172. &  9 & 70.8 & 65.2--79.5 & 60.1--91.1 \\
11 & 200. &  9 & 77.0 & 69.5--88.2 & 65.9--94.7 \\
12 & 237. &  9 & 78.2 & 68.7--81.6 & 63.8--85.1 \\
13 & 282. & 10 & 72.2 & 58.5--78.5 & 53.2--117. \\
14 & 385. & 10 & 48.4 & 39.3--55.5 & 37.1--78.2 \\
15 & 598. & 10 & 44.2 & 41.1--48.3 & 39.1--54.6 \\
\cutinhead{11 measurements per bin\tablenotemark{b}}
1  & 8.00 & 11 & 27.4 & 23.8--29.9 & 22.3--32.0 \\
2  & 52.7 & 11 & 33.4 & 30.8--39.9 & 29.5--51.2 \\
3  & 73.5 & 11 & 39.1 & 37.3--46.0 & 33.8--47.9 \\
4  & 94.0 & 12 & 48.2 & 45.0--52.0 & 42.7--55.2 \\
5  & 112. & 10 & 54.3 & 47.6--56.0 & 34.2--64.7 \\
6  & 133. & 12 & 67.4 & 56.3--74.0 & 46.3--80.6 \\
7  & 147. & 11 & 64.0 & 54.9--70.9 & 42.3--75.3 \\
8  & 172. & 11 & 71.1 & 65.9--80.5 & 62.4--87.7 \\
9  & 207. & 11 & 78.3 & 78.0--85.9 & 69.9--96.1 \\
10 & 257. & 11 & 77.1 & 66.2--84.2 & 62.2--115. \\
11 & 326. & 12 & 55.8 & 51.6--61.0 & 45.6--74.5 \\
12 & 592. & 12 & 42.2 & 39.2--46.9 & 39.1--49.0 \\
\cutinhead{13 measurements per bin\tablenotemark{b}}
1  & 10.4 & 14 & 28.8 & 26.7--31.4 & 23.1--32.1 \\
2  & 58.2 & 13 & 38.3 & 35.0--47.0 & 31.0--55.0 \\
3  & 86.6 & 14 & 45.0 & 41.9--47.0 & 35.9--48.4 \\
4  & 108. & 14 & 54.4 & 50.4--55.8 & 42.2--61.0 \\
5  & 133. & 14 & 58.7 & 55.3--72.3 & 43.6--78.5 \\
6  & 153. & 13 & 65.6 & 64.5--70.7 & 61.1--74.2 \\
7  & 184. & 13 & 78.1 & 72.1--86.0 & 67.3--91.6 \\
8  & 243. & 14 & 79.5 & 78.1--83.7 & 65.3--99.5 \\
9  & 317. & 14 & 57.7 & 53.1--70.1 & 49.4--75.7 \\
10 & 592. & 12 & 42.2 & 39.2--46.9 & 39.1--49.0 \\
\cutinhead{15 measurements per bin\tablenotemark{b}}
1 & 10.4 & 14 & 28.8 & 26.7--31.4 & 23.1--32.1 \\
2 & 61.4 & 16 & 37.5 & 34.1--42.6 & 31.4--52.1 \\
3 & 91.7 & 15 & 47.0 & 45.1--48.7 & 43.1--52.1 \\
4 & 114. & 15 & 56.0 & 54.9--60.0 & 53.0--70.1 \\
5 & 139. & 15 & 60.7 & 53.3--65.6 & 39.4--74.1 \\
6 & 171. & 14 & 72.8 & 67.1--78.4 & 65.4--83.7 \\
7 & 216. & 16 & 78.3 & 73.7--81.5 & 66.5--87.4 \\
8 & 300. & 15 & 70.7 & 59.2--78.0 & 52.8--114. \\
9 & 550. & 15 & 42.6 & 39.2--48.0 & 39.1--51.1 \\
\enddata
\tablenotetext{a}{Median of $\ell_{\rm e}$ values of measurements in the bin.}
\tablenotetext{b}{The lower cutoffs on the integral of the likelihood function
  are 3, 0.1, 0.1, and 0.1 $\mu$K for 9, 11, 13, and 15 measurements per bin.}
\end{deluxetable}

\clearpage

\centerline{\bf FIGURE CAPTIONS}

\begin{itemize}

\item[Fig. 1. -- ] {CMB anisotropy bandtemperature predictions and observational 
results, as a function of multipole $\ell$. Colored hatched regions are
adiabatic CDM model predictions for what would 
be seen by a series of ideal, Kronecker-delta window-function, experiments.
(That is, the model predictions do not account for the experiment window
functions.)  These
are for baryonic density parameter $\Omega_B = 0.0125 h^{-2}$ (where the
Hubble constant $H_0 = 100 h$ km s$^{-1}$ Mpc$^{-1}$) and are normalized
to the $\pm$1 $\sigma$ range allowed by the DMR measurements (G\'orski et al.
1998; Stompor 1997). Green is a flat-$\Lambda$ model with nonrelativistic
matter density parameter $\Omega_0 = 0.4$ and $h = 0.6$, red is an open
model with $\Omega_0 = 0.4$ and $h = 0.65$, and blue is fiducial CDM with
$\Omega_0 = 1$ and $h = 0.5$. ``Points" with different symbols represent 
observational results. Since most of the 
smaller-scale data points are derived assuming a flat bandpower CMB anisotropy
angular spectrum, which is more accurate for narrower (in $\ell$) window
functions, we have shown the observational results from the narrowest windows
available. Open symbols with inserted solid inverted triangles are 
2 $\sigma$ upper limits. There are 37 2 $\sigma$ upper limits but 17 of these 
lie above $\delta T_\ell = 120$ $\mu$K and so are not shown on the plot. (In
those cases for which a proper 2 $\sigma$ upper limit has not been quoted
by the observational group we have simply doubled the quoted 1 $\sigma$ 
upper error bar. Such approximate 2 $\sigma$ upper limits likely 
underestimate the true 2 $\sigma$ upper limits. Note that the two upper 
limits with $\delta T_\ell < 58 \mu$K at $\ell \sim 400$ fall in to this
category.)  Detections have $\pm$1 $\sigma$ vertical error bars. There 
are 142 detections but 2 of them (SK94Q9 and SK95C15) lie off the top of the
plot. Horizontal ``error bars" represent the width of the corresponding window 
functions. These data will eventually be available at   
www.phys.ksu.edu/$\sim$tarun/CMBwindows/wincomb/wincomb\_tf.html. 
The data shown are from the DMR galactic frame maps ignoring
the Galactic emission correction (G\'orski 1997, open octagons with $\ell 
\leq 20$); FIRS (Bond 1995, solid pentagon); Tenerife (Guti\'errez et al. 
2000, open five-point star); Python I--III and V (Rocha et al. 1999 and 
T. Souradeep, private communication 2000, open six-point stars); BOOMERanG
1997 and 1998 (Mauskopf et al. 2000; de Bernardis et al. 2000, open four-point
stars); IAC--Bartol (Femen\'{\i}a et al. 1998, open four-point diamond); 
Saskatoon 1993--95 (Netterfield et al. 1997, open squares); UCSB South Pole 
1994 (Ganga et al. 1997a, solid circles); BAM (Tucker et al. 1997, open circle);
MAT 1997 and 1998 (Torbet et al. 1999; Miller et al. 1999, open pentagons); 
MAXIMA-1 (Hanany et al. 2000, skeletal stars); QMAP 1 and 2 (Devlin et al.
1998; Herbig et al. 1998, solid pentagons); ARGO Hercules (Ratra et al. 1999a,
solid square); Jodrell Bank--IAC (Dicker et al. 1999; Harrison et al. 2000,
solid hexagons); Viper (Peterson et al. 2000, open seven-point stars); MAX3--5, 
(J. Gundersen, private communication, 1995;  S. Tanaka private communication
1995; Ganga et al. 1998, open hexagons); MSAM combined (Wilson et al. 2000, 
solid four-sided
diamonds); CAT 1 and 2 (G. Rocha, private communication, 1997; Baker et al.
1999, open four-sided diamonds); OVRO (Leitch et al. 2000, open five-point
star); and White Dish (Ratra et al. 1998, open pentagon).}

\item[Fig. 2. -- ]{CMB anisotropy bandtemperature predictions (colored hatched 
regions --- models are described in the caption of Figure 1), binned 
weighted-mean observational detection data for all 142 measurements (solid
black points connected by a solid black line are the central values and the 
other four solid black lines are the $\pm$1 $\sigma$ and $\pm$2 $\sigma$
limits), 10 times the number of standard deviations the weighted-mean result 
deviates from what is expected on the basis of Gaussianity of the CMB 
anisotropy (dashed line), and observational 2 $\sigma$ upper limits, all as a 
function of multipole $\ell$. Note that the model predictions here (and in  
subsequent figures) have not been processed in the same manner as the 
observational data. This 
is because the window functions of some experiments are not available.} 

\item[Fig. 3. -- ]{CMB anisotropy bandtemperature predictions (colored hatched 
regions --- models are described in the caption of Figure 1), binned 
median-statistics observational detection data for all 142 measurements (solid
black points connected by a solid black line are the central values and the 
other four solid black lines are the $\pm$1 $\sigma$ and $\pm$2 $\sigma$
limits), and observational 2 $\sigma$ upper limits, all as a function of 
multipole $\ell$.}

\item[Fig. 4. -- ]{CMB anisotropy bandtemperature predictions (colored hatched 
regions --- models are described in the caption of Figure 1), binned 
weighted-mean observational detection data for the culled data with 135 
measurements (solid black points connected by a solid black line are the 
central values and the other four solid black lines are the $\pm$1 $\sigma$ 
and $\pm$2 $\sigma$ limits), 10 times the number of standard deviations the 
weighted-mean result deviates from what is expected on the basis of Gaussianity 
of the CMB anisotropy (dashed line), and observational 2 $\sigma$ upper limits, 
all as a function of multipole $\ell$.}

\item[Fig. 5. -- ]{CMB anisotropy bandtemperature predictions (colored hatched 
regions --- models are described in the caption of Figure 1), binned 
median-statistics observational detection data for the culled data with 135 
measurements (solid
black points connected by a solid black line are the central values and the 
other four solid black lines are the $\pm$1 $\sigma$ and $\pm$2 $\sigma$
limits), and observational 2 $\sigma$ upper limits, all as a function of 
multipole $\ell$.}

\end{itemize}
\clearpage

\begin{figure}
\resizebox{\textwidth}{!}{\includegraphics{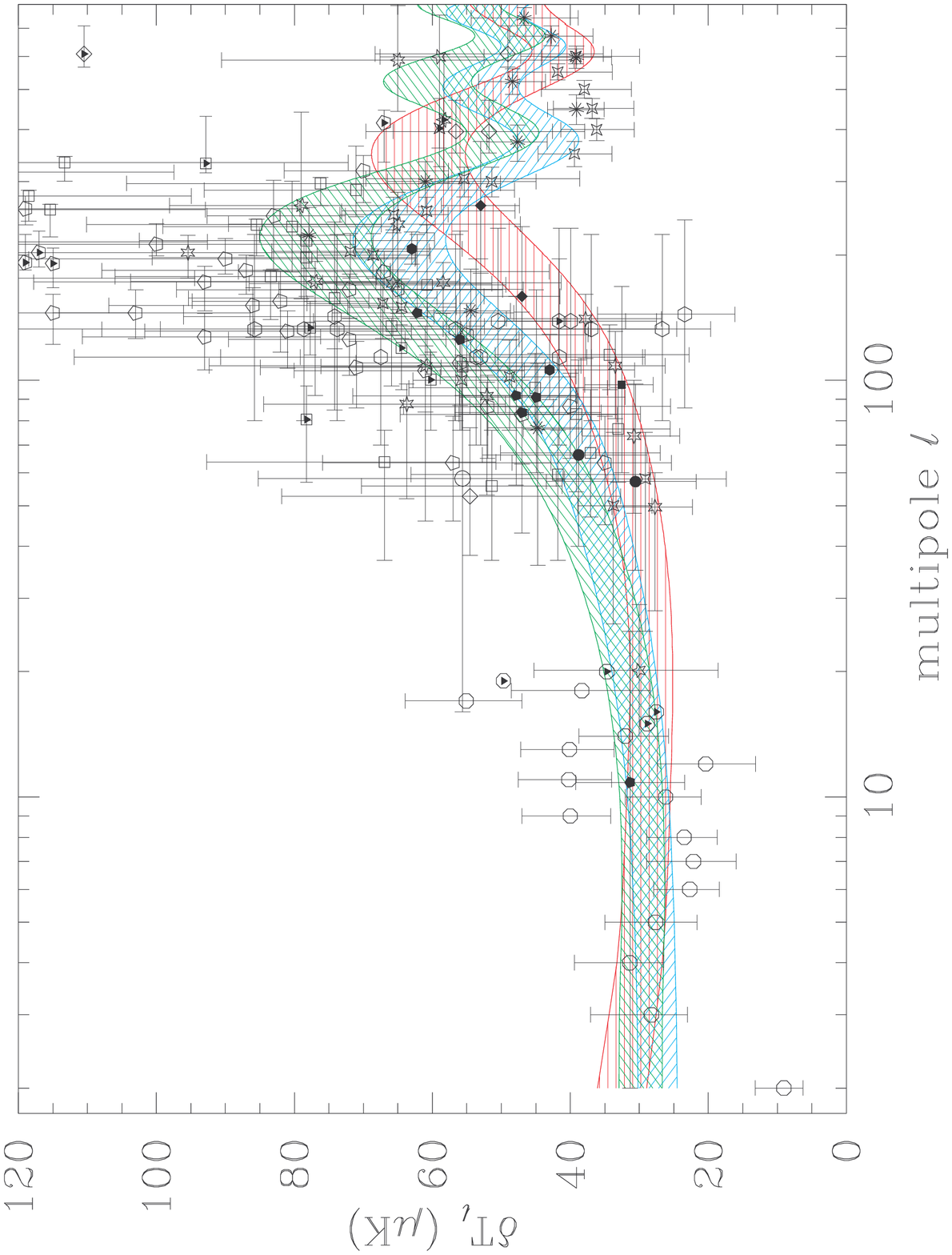}}
Figure 1
\end{figure}
\clearpage

\begin{figure}
\resizebox{\textwidth}{!}{\includegraphics{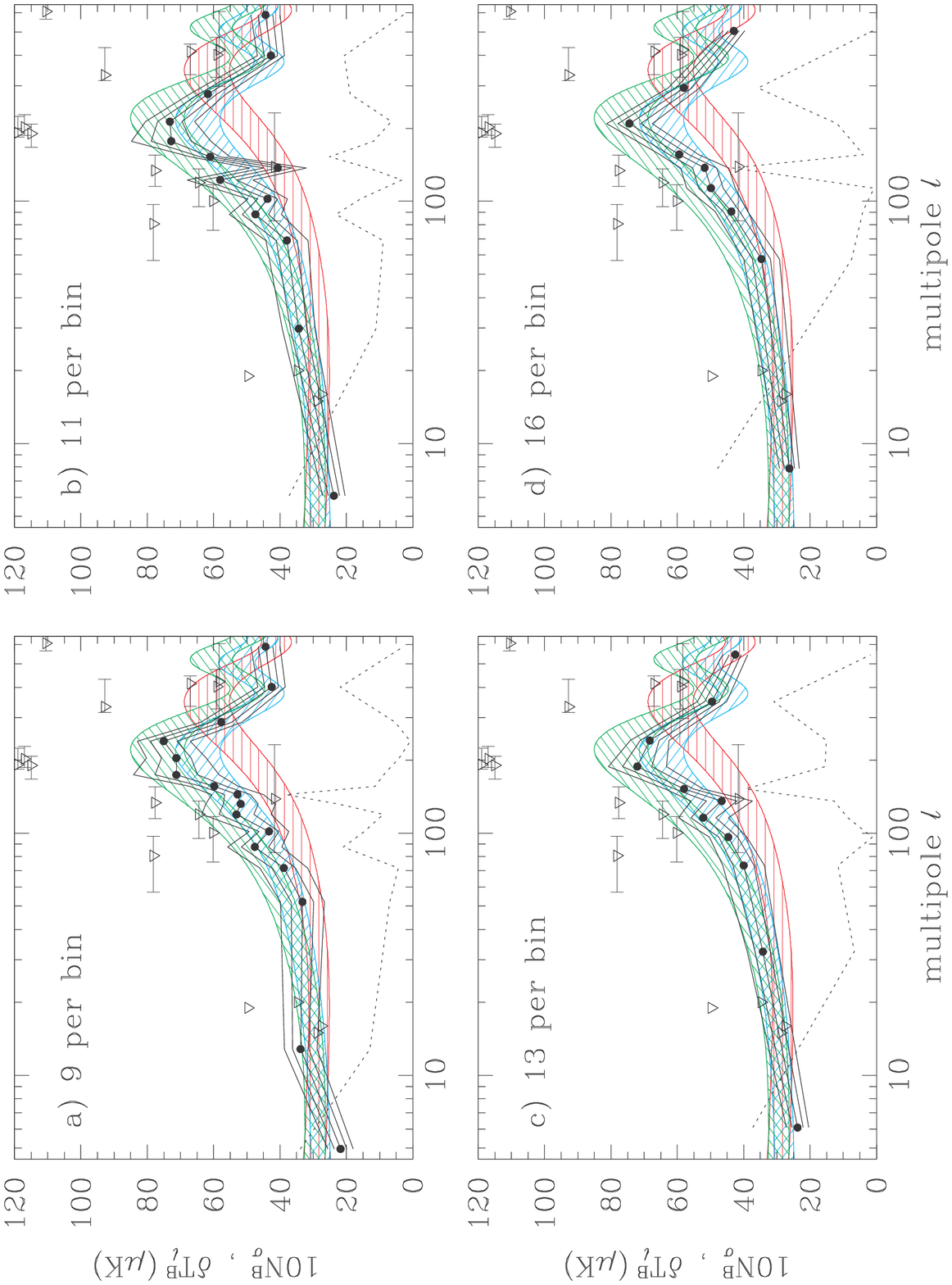}}
Figure 2
\end{figure}
\clearpage

\begin{figure}
\resizebox{\textwidth}{!}{\includegraphics{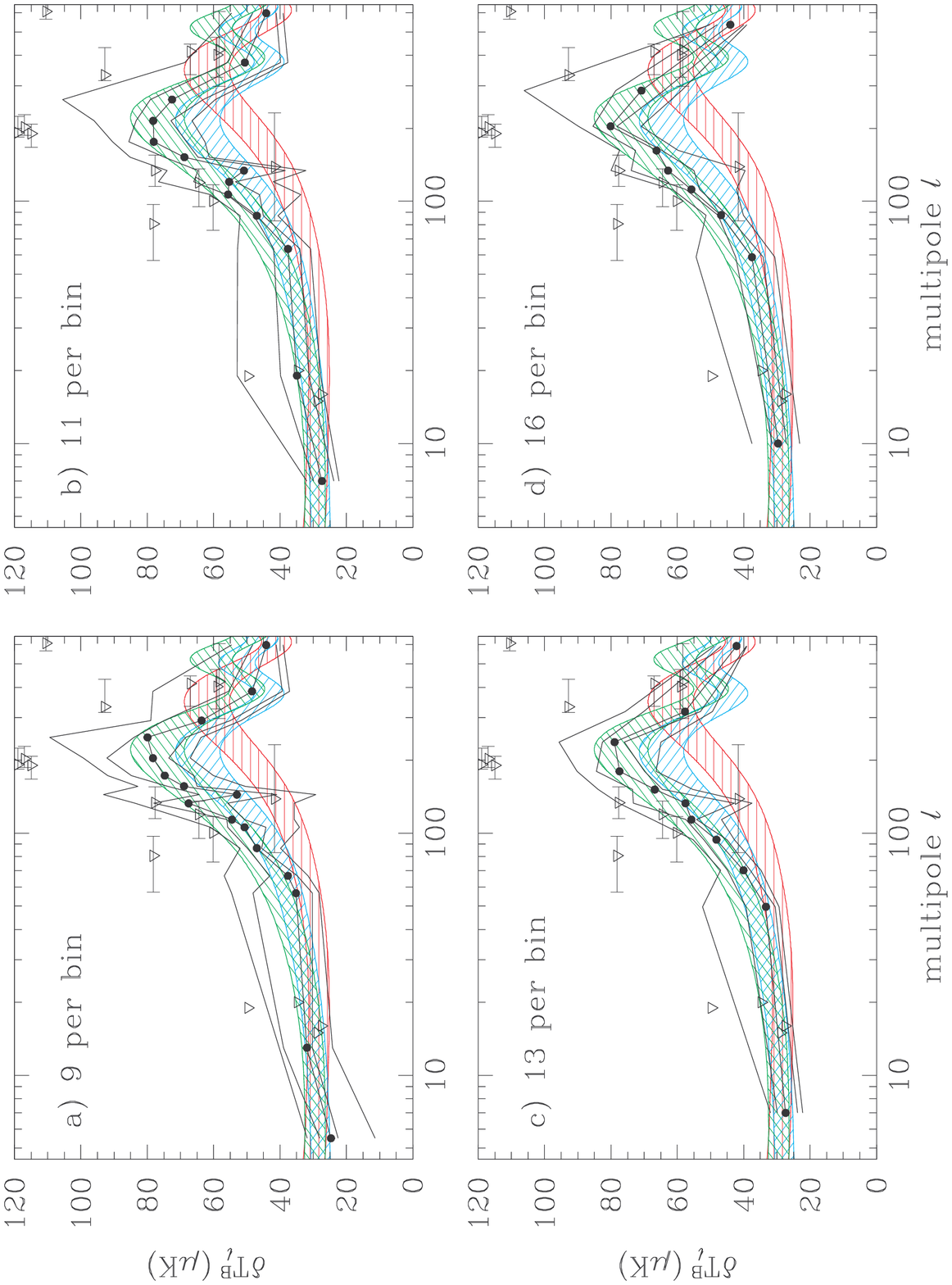}}
Figure 3
\end{figure}
\clearpage

\begin{figure}
\resizebox{\textwidth}{!}{\includegraphics{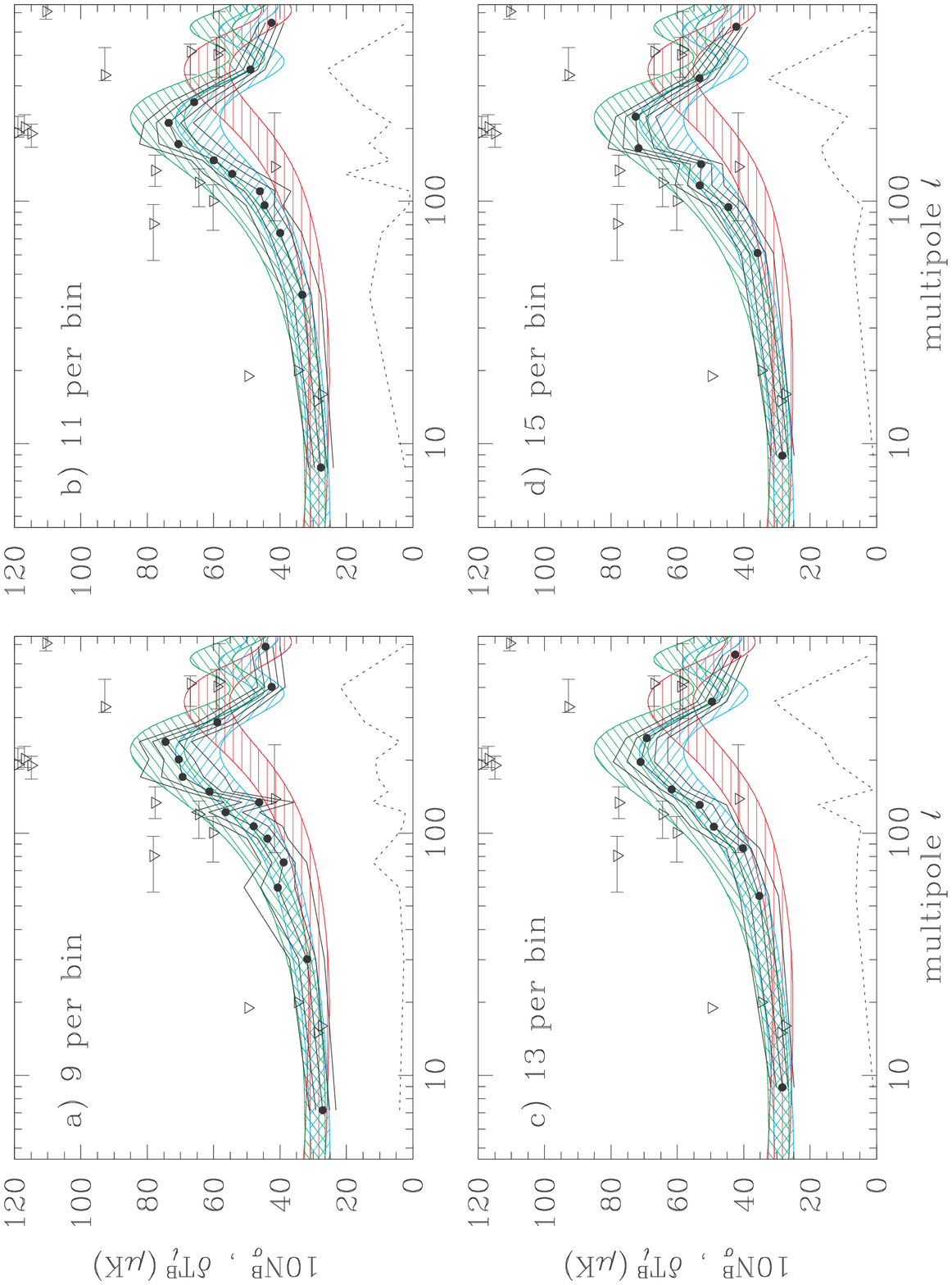}}
Figure 4
\end{figure}
\clearpage

\begin{figure}
\resizebox{\textwidth}{!}{\includegraphics{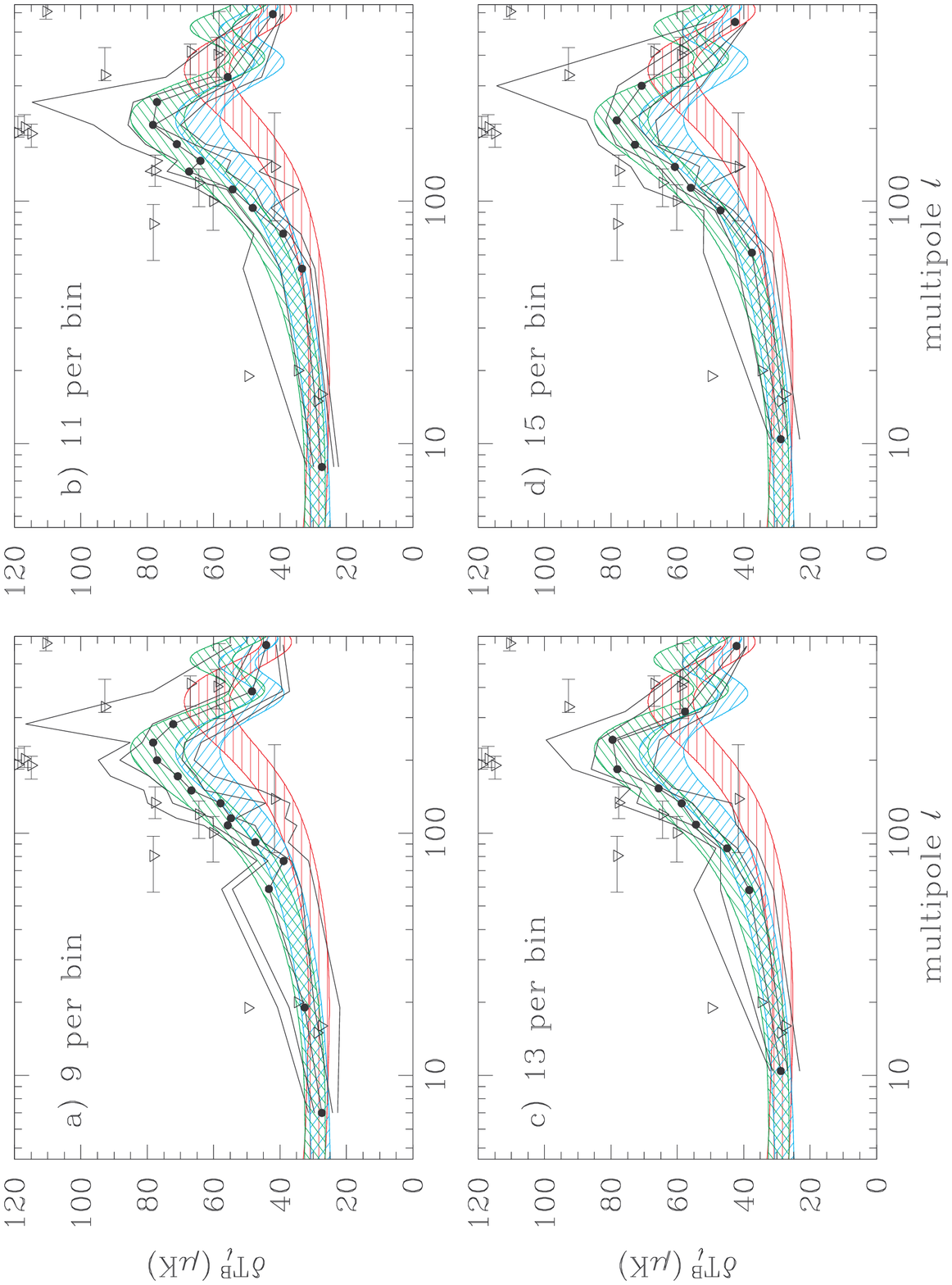}}
Figure 5
\end{figure}
\clearpage

\end{document}